\documentclass[aps,prstab,preprint,showpacs,groupedaddress]{revtex4}
\usepackage{graphics,epsfig}
\begin{document}
\def\be{\begin{equation}}
\def\ee{\end{equation}}
\def\bc{\begin{center}}
\def\ec{\end{center}}
\def\bea{\begin{eqnarray}}
\def\eea{\end{eqnarray}}
\draft
\title{Scale-free networks with tunable degree distribution exponents}
\author{H. Y. Lee, H. Y. Chan, and P. M. Hui}
\address{Department of Physics, The Chinese University of Hong
Kong,\\
New Territories, Shatin, Hong Kong, China}

\begin{abstract}

We propose and study a model of scale-free growing networks that
gives a degree distribution dominated by a power-law behavior with
a model-dependent, hence tunable, exponent. The model represents a
hybrid of the growing networks based on popularity-driven and
fitness-driven preferential attachments.  As the network grows, a
newly added node establishes $m$ new links to existing nodes with
a probability $p$ based on popularity of the existing nodes and a
probability $1-p$ based on fitness of the existing nodes. An
explicit form of the degree distribution $P(p,k)$ is derived
within a mean field approach. For reasonably large $k$, $P(p,k)
\sim k^{-\gamma(p)}{\cal F}(k,p)$, where the function ${\cal F}$
is dominated by the behavior of $1/\ln(k/m)$ for small values of
$p$ and becomes $k$-independent as $p \rightarrow 1$, and
$\gamma(p)$ is a model-dependent exponent. The degree distribution
and the exponent $\gamma(p)$ are found to be in good agreement
with results obtained by extensive numerical simulations.

\end{abstract}



\maketitle

Many complex systems, including social, biological, physical,
economic, and computer systems, can be studied by network models
in which the nodes represent the constituents and links or edges
represent the interactions between constituents
\cite{albert1,dorogovtsev1,mendesbook,wattsbook1,wattsbook2}.
Interesting findings in the statistics of real-world networks
reveal that classical random networks \cite{erdos,bollobas} do not
often represent the geometrical or topological structure of
real-world networks
\cite{albert2,huberman,broder,redner,jeong1,jeong2,albert3,liljeros,granovetter,newman2,berlow}
correctly.  Most noticeable of the properties observed are
so-called ``six degree of separation" \cite{milgram,watts} from
one node to any arbitrary node, and the highly clustering feature.
In particular, many networks show a power-law degree distribution
of the form $P(k)\sim k^{-\gamma}$, with the exponent $\gamma$
taking on values between $2$ to $3$.  This behavior has led to the
construction of models of scale-free growing networks. Barab\'asi
and Albert (BA) \cite{barabasi1} proposed a model in which a new
node is added in each turn and $m$ new links are established with
existing nodes with the probability of establishing a link being
proportional to the number of existing links of the nodes. This
preferential attachment is thus driven entirely by the popularity
of existing nodes. Detailed numerical simulations and analytic
analysis showed that $\gamma =3$ for the BA model. The idea of
incorporating preferential attachment in a growing network has led
to proposals of a considerable number of models of scale-free
networks \cite{krapivsky1,dorogovtsev2,albert4,krapivsky2}.
Alternatively, models with preferential attachment driven entirely
by fitness have also been proposed \cite{bianconi,ergun}.  In
these models, each node carries a randomly assigned fitness value
that gives a collective character of the node other than its
popularity, and the probability of establishing a new link to an
existing node is proportional to the product of fitness and the
number of existing links.  It was found that the degree
distribution $P(k) \sim k^{-\gamma}/\ln(k/m)$, with $\gamma =
2.255$ for the fitness-driven model \cite{bianconi}.

Many real-world networks such as paper citations in scientific
journals, the world-wide web, the internet, and the collaborative
networks of actors and actresses, exhibit a degree distribution
with a network-dependent exponent that takes on values close to
but below $3$.  It is therefore interesting to construct and
analyze models with a tunable degree-distribution exponent.  In
the present work, we propose and study a model representing a
hybrid of the growing network models based on popularity-driven
and fitness-driven preferential attachments of new links.  As the
network grows, a newly added node has a probability $p$ of being
popularity-driven and a probability $1-p$ of being fitness-driven
in establishing new links.  Thus the resulting network consists of
a mixture of two types of nodes, with a fraction $p$ establishing
new links based entirely on popularity consideration.  The model
reflects the fact that not every one, taking the nodes as agents
in a population, prefers to follow the popular persons, but
instead may prefer to establish relationships with others based on
characters other than the popularity of the agents.  Our model
thus incorporates the inhomogeneous nature of many real-world
networks in which not all the nodes are identical. We report
results of extensive numerical simulations on the degree
distribution for networks of size $10^{7}$ nodes, and compare
numerical results with an analytic expression derived via a mean
field approach. For reasonably large $k$, the degree distribution
$P(p,k)$ follows the form $k^{-\gamma(p)}{\cal F}(k,p)$, where
${\cal F}(k,p)$ is dominated by the behavior of $1/\ln(k/m)$ for
small $p$ and becomes $k$-independent for $p \rightarrow 1$. The
exponent $\gamma(p)$ can be extracted numerically and results are
found to be in good agreement with that of the mean field theory.

Our model is defined as follows.  Initially a fully connected
network of $m_{0}$ nodes is constructed, with $m_{0}$ typically of
order unity.  In this work, we use $m_{0}=5$.  The network grows
with one new node being added to the existing network at a time.
Each newly added node establishes a number of $m$ new links to
existing nodes. With probability $p$, the new node establishes
links by preferential attachments based on popularity of the
existing nodes \cite{barabasi1}, i.e., the probability that an
existing node $i$ is connected is proportional to the degree or
the number of links $k_{i}(t)$ that node $i$ carries.  With
probability $1-p$, the new node establishes links by preferential
attachments based on fitness of the existing nodes
\cite{bianconi}, i.e., the probability that an existing node $i$
is connected is proportional to the product $k_{i}(t) \eta_{i}$,
where the fitness $\eta$ of a node is a randomly assigned value in
the interval $0 < \eta < 1$ associated with the node when it is
introduced into the network. For $p=1$ ($p=0$), the present model
reduces to the popularity-driven BA \cite{barabasi1}
(fitness-driven \cite{bianconi}) model.

Detailed numerical simulations have been carried out for our model
with networks of size up to $10^{7}$ nodes.  Each newly added node
establishes $m=5$ new links. Figure 1 shows a typical degree
distribution function on a log-log scale for the case of $p=0.5$,
i.e., a new node randomly chooses to follow the popularity-driven
or fitness-driven rules in establishing new links, and for $p=0.2$
(see inset). The large size of the networks used in this study
makes the comparison with analytic results and the extraction of
the exponent in the degree distribution easier.  The data shown in
Fig.1 represent an average over 10 different realizations of
networks of the same size.  To explore the functional form of the
degree distribution and to extract possible exponent, we need some
guidance from analytic treatment.

The model can be treated analytically by a mean field approach
\cite{albert1,dorogovtsev1,bianconi}. For sufficiently long time,
the connectivity $k_{i}(t)$ of the $i$-th node with fitness
$\eta_{i}$ evolves according to the following continuous time
evolution equation
\begin{equation}
\frac{\partial k_{i}}{\partial t} = m p \frac{k_{i}(t)}{\sum_{j}
k_{j}} + m (1-p) \frac{\eta_{i}k_{i}(t)}{\sum_{j} \eta_{j}k_{j}},
\end{equation}
where the first and second terms describe the increment in $k_{i}$
due to popularity and fitness, respectively.  Note that $\sum_{j}
k_{j} = 2mt$, with the factor of $2$ coming from the undirected
nature of the links. We assume that $k_{i}$ takes on the form
\begin{equation}
k_{i}(t,t_{0}) = m \left(
\frac{t}{t_{0}}\right)^{\beta(\eta_{i},p)},
\end{equation}
where $t_{0}$ is the time at which the $i$-th node was introduced
into the network.  Since $k_{i}$ can at most be increased by one
at each time step, it cannot grow faster than $t$, thus implying
$0 < \beta(\eta_{i},p) < 1$.

The fitness $\eta_{i}$ is, in general, chosen from a distribution
$\rho(\eta)$. The average of the sum $\sum_{j} \eta_{j} k_{j}$
over $\rho(\eta)$ can be evaluated by
\begin{eqnarray}
\langle \sum_{j} \eta_{j} k_{j} \rangle & = & \int \eta_{j}
\rho(\eta_{j}) \int_{1}^{t} k_{j}(t,t_{0}) dt_{0} d \eta_{j}
\nonumber \\
& = & \int \eta_{j} \rho(\eta_{j})
\frac{m(t-t^{\beta(\eta_{j},p)})}{1 - \beta(\eta_{j},p)}
d\eta_{j}.
\end{eqnarray}
For large $t$, since $t^{\beta}/t \rightarrow 0$, the contribution
from the term $t^{\beta(\eta_{j},p)}$ becomes negligible and we
have
\begin{equation}
\langle \sum_{j} \eta_{j} k_{j} \rangle = C(p) m t,
\end{equation}
where
\begin{equation}
C(p) = \int \rho(\eta_{j}) \frac{\eta_{j}}{1 - \beta(\eta_{j},p)}
d\eta_{j}.
\end{equation}
Substituting Eqs.(2) and (4) into Eq.(1), we obtain
\begin{equation}
\beta(\eta_{i},p) = \frac{p}{2} + \frac{1-p}{C(p)} \eta_{i}.
\end{equation}
From now on, we drop the subscripts $i$ and $j$ for brevity.
Equation (6) and Eq.(5) can be solved self-consistently for
$C(p)$.  For $\rho(\eta)$ being a uniform distribution between $0$
and $1$, i.e., $\rho(\eta) = 1$ for $\eta \in [0,1]$, $C(p)$
satisfies
\begin{equation}
C(p) = \int_{0}^{1} \frac{\eta}{(1- \frac{p}{2}) -
\frac{(1-p)\eta}{C(p)}} d\eta,
\end{equation}
for which the integral can be performed to obtain the
self-consistent equation
\begin{equation}
\frac{1}{C(p)} = \frac{1-p/2}{1-p} \left(1 -
e^{-\frac{2(1-p)}{C(p)}} \right).
\end{equation}
Equation (8) can be solved numerically for $C(p)$, for given value
of $p$.  It is found that $C(p) \in [1,1.255]$ for $p \in [0,1]$.
Note that for $p=1$, $\beta = 1/2$ (see Eq.(6)) independent of
$\eta$ as in the BA popularity-driven model \cite{barabasi1}.  For
$p=0$, $C(0)=1.255$ and $\beta(\eta,0) = \eta/1.255$, as in the
purely fitness-driven model \cite{bianconi}.

To proceed, the cumulative probability distribution function (CDF)
${\cal P}_{\eta}(p,k_{i}>k)$ for a particular fitness $\eta$ and
given $p$ can be found by noting that a degree higher than some
value $k$ for a node corresponds to a cutoff in time before which
the node must have been introduced into the network, i.e.,
\begin{eqnarray}
{\cal P}_{\eta}(p,k_{i}>k) & = & {\cal P}_{\eta}(t_{0} < t \left(
\frac{m}{k} \right)^{1/\beta(\eta,p)})  \nonumber \\
& = & \left( \frac{m}{k} \right)^{1/\beta(\eta,p)}.
\end{eqnarray}
Note that a pre-factor of $t/(m_{0}+t)$, which approaches unity
for sufficiently long time, has been ignored in Eq.(9). To obtain
the CDF for the whole network ${\cal P}_(p,k)$, an average is
taken over a uniform distribution of $\eta$, thus
\begin{equation}
{\cal P}(p,k) = \int_{0}^{1} \left( \frac{m}{k}
\right)^{1/\beta(\eta,p)} d\eta,
\end{equation}
with $\beta(\eta,p)$
given by Eq.(6). The probability distribution function (PDF) of
degrees in the network $P(p,k)$ is related to ${\cal P}(p,k)$
through
\begin{equation}
P(p,k) = - \frac{\partial}{\partial k} {\cal P}(p,k).
\end{equation}
Using Eq.(10) and by making the substitution $x=\ln(k/m)/(p/2 +
(1-p)\eta/C(p))$, Eq.(11) gives
\begin{equation}
P(p,k) = \frac{C(p)}{k(1-p)} \int_{x_{0}}^{x_{1}} \frac{e^{-x}}{x}
dx,
\end{equation}
where the lower limit of the integral $x_{0} = \alpha_{0} ln(k/m)$
and the upper limit $x_{1} = \alpha_{1} \ln(k/m)$, with
\begin{equation}
\alpha_{0} = \frac{1}{p/2 + (1-p)/C(p)}
\end{equation}
and
\begin{equation}
\alpha_{1} = \frac{2}{p}.
\end{equation}

We aim at getting the functional form of $P(p,k)$ and in
particular the dependence on $k$ at given $p$. The integral in
Eq.(12) can be carried out by parts, and only the ``surface" term
survives in the limit of $\ln(k/m) \gg 1$. It follows from the
fact that $1/x$ takes on the maximum value of $x_{0}^{-1}$ for $x$
in the interval $[x_{0},x_{1}]$, and thus the integral
$\int_{x_{0}}^{x_{1}} (e^{-x}/x^{2}) dx \leq x_{0}^{-1}
\int_{x_{0}}^{x_{1}} (e^{-x}/x) dx$, with $x_{0}^{-1} \ll 1$ for
large $k$.  Equation (12) thus gives
\begin{equation}
P(p,k) = \frac{C}{m(1-p)}\cdot \frac{(\frac{k}{m})^{-(1 +
\alpha_{0})}}{\alpha_{0}\ln(k/m)} \left(1 -
\frac{\alpha_{0}}{\alpha_{1}} \left(\frac{k}{m}
\right)^{-\alpha_{1}+\alpha_{0}} \right).
\end{equation}
Equation (15) is the main result of the mean field treatment.  It
gives the explicit $k$-dependence of the degree distribution
function. It is worth noting that Eq.(15) gives the correct
results in both limits of $p \rightarrow 0$ and $p \rightarrow 1$.
For purely fitness-driven model, $P(0,k) =
(k/m)^{-2.255}/(mln(k/m))$ \cite{bianconi}.  For purely
popularity-driven model, the $\ln(k/m)$ term in the denominator
can be shown to be cancelled by the terms in the parentheses,
giving $P(1,k) = 2m^{2}k^{-3}$ \cite{barabasi1}.

Results obtained from the mean field theory can be compared with
numerical results. The solid lines in Fig.1 show the degree
distribution for $p=0.5$ and $p=0.2$ (inset) using Eq.(12).
Excellent agreements are found between mean field and numerical
results. Equation (15) also suggests a functional form for
$P(p,k)$.  For reasonably large $k$, e.g., $k \sim 10^{2}$ or
above, the second term in the parentheses is small compared to
unity, especially for small values of $p$.  Hence, Eq.(15)
suggests that $P(p,k) \sim k^{-\gamma(p)}/\ln(k/m)$, where
\begin{equation}
\gamma(p) = 1+ \alpha_{0} = 1 + \frac{1}{(p/2 + (1-p)/C(p))}
\end{equation}
with $C(p)$ given by Eq.(8) within mean field theory. Numerically,
we fit the degree distribution function to the form
$k^{-\gamma(p)}/\ln(k/m)$ and extract the exponent $\gamma(p)$
directly from results of numerical simulations for each value of
$p$ in steps of $0.1$. However, as $p \rightarrow 1$, one should
be more careful in handling Eq.(15) in numerical extraction of the
exponent as the prefactor $1/(1-p)$ diverges and the $\ln(k/m)$
term becomes unimportant due to cancellation effect from the terms
in the parentheses.  In this case, it is more convenient to work
from the $p=1$ limit and extract a functional form from Eq.(15)
that is valid for $p \rightarrow 1$. The result is $P(p,k) \sim
(p/2 + (1-p)/C)(k/m)^{-\gamma(p)}/mp^{2}$, where $\gamma(p)$ is
again given by Eq.(16). This form is used to extract the exponent
$\gamma(p)$ for $p=0.9$ and $1.0$. For $p \leq 0.8$, using either
functional form extracts the same value of $\gamma$. Figure 2
shows the exponent $\gamma(p)$ numerically extracted from
simulations, together with the analytic result. The two sets of
results are found to be in good agreement.

In summary, we proposed and studied a model in which the nodes are
inhomogeneous.  The model combines popularity-driven and
fitness-driven preferential attachments in growing networks.
Extensive numerical simulations were carried out and a mean field
theory was developed.  The degree distribution function shows a
predominant power-law behavior.  The exponent takes on a
model-dependent, hence tunable, value depending on the
concentration $p$ of nodes for which the links are established by
a popularity-driven mechanism. The exponent $\gamma(p)$ takes on
values between $2.255$ and $3$, which lie within the range of
values of $\gamma$ observed in many real-world networks.  Analytic
expressions for the degree distribution function and the exponent
$\gamma(p)$ were derived. Results of mean field theory were found
to be in good agreement with results obtained by numerical
simulations.

\acknowledgments{We wish to thank Oliver K.H. Chung for useful
discussions on numerical methods on setting up large networks
efficiently, and Prof. DaFang Zheng for useful discussions on
fitness-driven growing networks.}

\newpage
\centerline{\bf FIGURE CAPTIONS}

\bigskip
\noindent Figure 1: The degree distribution function $P(p,k)$ as a
function of $k$ on a log-log scale for $p=0.5$ and $p=0.2$
(inset). The symbols give numerical results averaging over 10
different realizations of networks of size of $10^{7}$ nodes.  The
lines give the analytic results within the mean field theory.

\bigskip
\noindent Figure 2: The exponent $\gamma(p)$ characterizing the
degree distribution function as obtained by fitting to numerical
simulation results (symbols) for networks of size $10^{7}$ nodes
for values of $p$ from $p=0$ to $p=1$ in steps of $0.1$, and by
the mean field theory (dotted line) given by $\gamma(p) = 1 +
1/(p/2 + (1-p)/C(p))$ with $C(p)$ given by Eq.(8).

\newpage

\newpage
\begin{figure}
\epsfig{figure=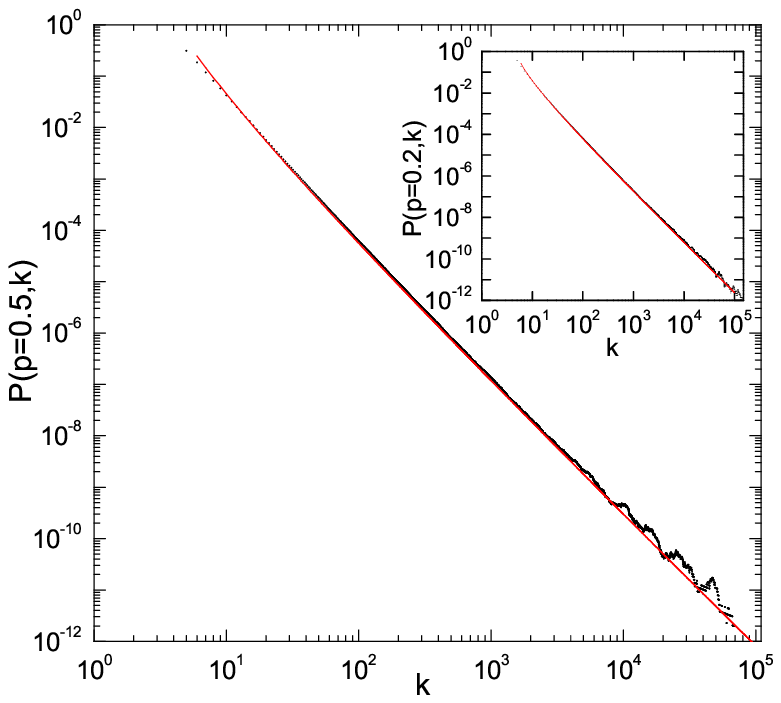,width=\linewidth} \label{figure1}
\end{figure}

\newpage
\begin{figure}
\epsfig{figure=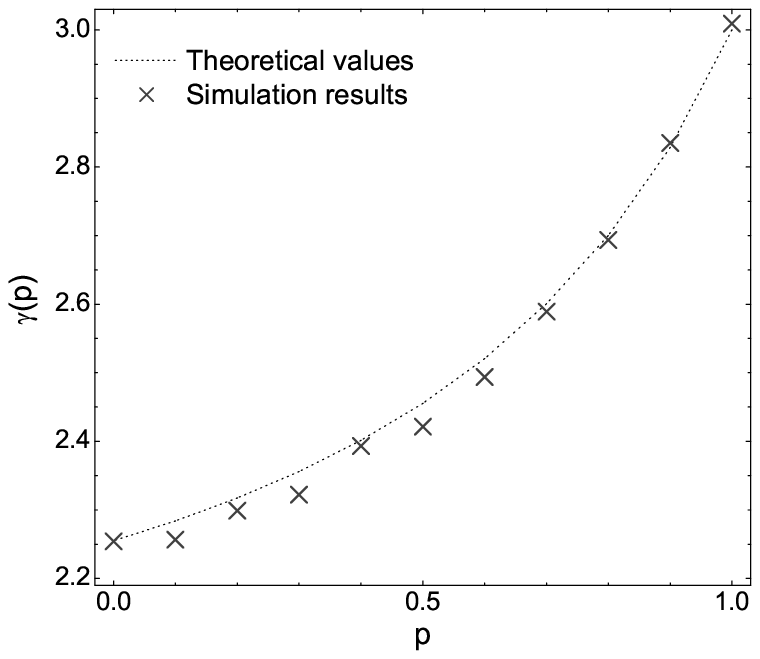,width=\linewidth} \label{figure2}
\end{figure}

\end{document}